\documentclass[amsmath,amssymb,floatfix,lengthcheck,superscriptaddress]{revtex4-1}
\usepackage[T1]{fontenc}
\usepackage{graphicx}
\usepackage[centering,hmargin=20mm,tmargin=29mm,bmargin=25mm]{geometry}
\usepackage{multirow}
\usepackage{newtxtext}
\usepackage[cmintegrals]{newtxmath} 
\usepackage{mathtools}
\usepackage{latexsym}
\usepackage{bm}
\usepackage{lineno}
\usepackage{lipsum}
\usepackage{soul}
\usepackage{tabularx}
\usepackage{xcolor}
\usepackage[pdfencoding=auto]{hyperref}
\usepackage{cleveref}
\hypersetup{colorlinks,
	linkcolor={blue!75!black!80!yellow},
	citecolor={blue!75!black!80!yellow},
	urlcolor={blue!75!black!80!yellow}
}

\makeatletter
\renewcommand\@make@capt@title[2]{%
	\@ifx@empty\float@link{\@firstofone}{\expandafter\href\expandafter{\float@link}}%
	\sffamily{\textbf{#1}}\@caption@fignum@sep#2
}%

\makeatother

\newcommand{\romone}{\uppercase\expandafter{\romannumeral1}}
\newcommand{\romtwo}{\uppercase\expandafter{\romannumeral2}}
\newcommand{\romthree}{\uppercase\expandafter{\romannumeral3}}

\begin{document}
\title{\emph{Ab initio} self-trapped excitons}

\author{Yunfei Bai}
\affiliation{Beijing National Laboratory for Condensed Matter Physics and Institute of Physics, Chinese Academy of Sciences, Beijing 100190, China}
\affiliation{School of Physical Sciences, University of Chinese Academy of Sciences, Beijing 100190, China}

\author{Yaxian Wang}
\email[Email address: yaxianw@iphy.ac.cn;]{}
\affiliation{Beijing National Laboratory for Condensed Matter Physics and Institute of Physics, Chinese Academy of Sciences, Beijing 100190, China}

\author{Sheng Meng}
\email[Email address: smeng@iphy.ac.cn;]{}
\affiliation{Beijing National Laboratory for Condensed Matter Physics and Institute of Physics, Chinese Academy of Sciences, Beijing 100190, China}
\affiliation{School of Physical Sciences, University of Chinese Academy of Sciences, Beijing 100190, China}
\affiliation{Songshan Lake Materials Laboratory, Dongguan, Guangdong 523808, China}

\begin{abstract}
We propose a new formalism and an effective computational framework to study self-trapped excitons (STE) in insulators and semiconductors from first principles. 
Using the many-body Bethe-Salpeter equation in combination with perturbation theory, we are able to obtain the mode- and momentum-resolved exciton-phonon coupling matrix element in a perturbative scheme, and explicitly solve the real space localization of the electron (hole), as well as the lattice distortion. 
Further, this method allows to compute the STE potential energy surface and evaluate the STE formation energy and Stokes shift.
We demonstrate our approach using two-dimensional magnetic semiconductor chromium trihalides and a wide-gap insulator BeO, the latter of which features dark excitons, and make predictions of their Stokes shift and coherent phonon generation which we hope to spark future experiments such as photoluminescence and transient absorption studies.

\end{abstract}
\date{\today}
\maketitle
Excitons, a quasiparticle composed of correlated electron-hole pairs, embrace rich many-body physics and play a crucial role in semiconductor science and technology.
As a fundamental quasiparticle, excitons can interact with the lattice degree of freedom, or phonons. 
Exciton-phonon interaction gives rise to a wide range of emergent phenomena such as phonon-induced exciton linewidth~\cite{Selig2016_excitonic_nc}, phonon side bands~\cite{Li2019_emerging_nc}, phonon replica~\cite{christiansen2017_phonon_prl}, and valley polarized exciton relaxation~\cite{chen2022_first_prr}, etc. 
Apart from these effects, strong exciton-phonon interaction can lead to a self-trapped exciton (STE), i.e. an exciton trapped on its own lattice distortion field
~\cite{Tan2022_selftrapped_nanoscale,Edamatsu1993transient,yang2021ultrafast,xu2021toward}, which can dramatically influence luminescence, energy transport, and formation of lattice defects in crystals. 
The STE is usually signaled by photoluminescence with a large and broadened Stokes shift, and coherent phonon generation which can be observed through transient absorption (TA) spectroscopy. 
STEs have been observed in various materials, including ionic crystal alkali halides studied extensively in the last century~\cite{song2013self}, two-dimensional (2D) magnetic chromium trihalides which have garnered great research interests recently~\cite{Seyler2017_ligand_np,Dillon1966_magneto_jpcs,grant1968optical,Pollini1970_intrinsic_pssb,Bermudez1979_spectroscopic_jpcs,Nosenzo1984_effect_prb}, the promising photovoltaic material CsPbBr$_3$~\cite{Tao2021_momentarily_nc}, and hybrid organic perovskites~\cite{Zhang2020_establishing_nc}.
Manipulating STE formation and their properties can further inspire new device concepts.
Their broad emission band, high photoluminescence quantum efficiency, and large tunability show promises for optical device applications such as light-emitting diodes~\cite{Wang2020_colloidal_nanolett,Chen2020_efficient_nphoto}, luminescent solar
concentrators~\cite{Meinardi2017}, and efficient photovoltaic cells~\cite{Zhao2013_transparent_aem}.

\begin{figure}[!ht]
    \centering
    \includegraphics[width=\linewidth]{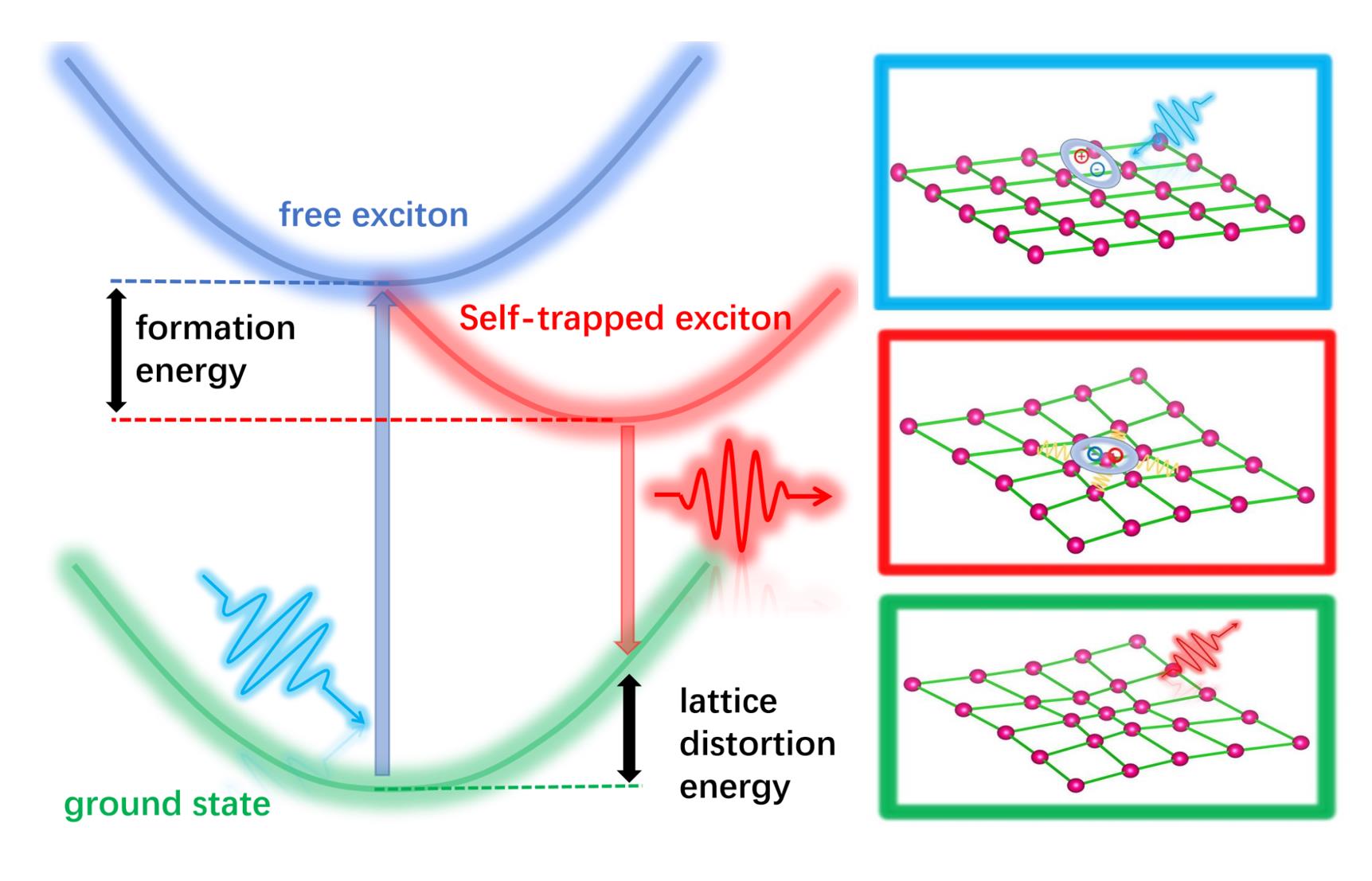}
    \caption{\textbf{Schematics showing the Franck-Condon picture of the STE formation. }
    The green, blue, and red curves in the left panel correspond to the potential energy surfaces (PESs) of the ground state, the free exciton state, and the STE state, respectively.
    The right panel illustrates interactions involved during the formation and decay of STE.
    Top: A semiconducting or insulating material is optically excited, generating a free exciton.
    Middle: The exciton is trapped by a localized lattice distortion, forming an STE state.
    Bottom: the STE decays with a lower-energy photon emitted, causing a Stokes shift as part of its energy is released to the lattice.}
    \label{fig:STE-schematics}
\end{figure}

Microscopically, the formation of STE can be understood based on the Franck-Condon principle, as schematically illustrated in Fig.~\ref{fig:STE-schematics}.
When a system is excited with electrons and holes generated, an exciton can form due to the Coulomb interaction. 
Meanwhile, the minimum of the excited state potential energy surface (PES) can possibly shift with respect to its ground state equilibrium, resulting in a finite lattice distortion.
The energy lowered in this process is defined as the STE formation energy (Fig.~\ref{fig:STE-schematics}).
Such lattice distortion serves as an attractive center to pin the otherwise free exciton.
When the STE decays, a photon emitted will have a Stokes shift since part of the energy is released to the lattice.

The STE effect is well observed experimentally, however,
an \emph{ab initio} description of STEs that properly accounts for the
electron-hole correlation effect and exciton-phonon interaction with the full lattice degree of freedom is still lacking and remains a daunting task, with the major challenges as follows.
First, although the supercell technique for calculating polarons, i.e. a quasiparticle formed by an excess electron (hole) with lattice distortion, has been widely applied~\cite{PhysRevB.90.035204,C4CP03981E}, the computational cost of the $GW$ with Bethe-Salpeter equation (BSE) approach makes the STE calculation using supercells impractical. 
Second, previous studies mainly adopt constrained density functional theory (cDFT) for lattice relaxation to calculate the STE configuration, and then use the BSE approach with the distorted lattice to calculate the exciton binding energy thus determining the photoluminescence spectrum~\cite{Li2022_ultrafast_nanolett}.
The major drawback is that missing the excitonic effect in the first place may lead to unrealistic lattice distortion, especially for strongly-correlated Frenkel excitons.
Although a Green's functions approach was proposed to calculate atomic forces with the presence of excitons~\cite{PhysRevLett.90.076401}, it is computationally demanding and limited to simple systems and relatively small supercells, thus hindering its wide applications.

Herein, we propose an \emph{ab initio} framework that properly treats both the excitonic effect and its coupling to all phonon modes, based on the many-body GW-BSE method~\cite{PhysRevB.34.5390,PhysRevB.62.4927,PhysRevB.86.115409} in combination with density functional perturbation theory (DFPT)~\cite{PhysRevLett.49.1691}. 
After benchmarking the STE lattice configuration in Li$_2$O$_2$, we present calculations for 2D magnetic chromium trihalides, using CrBr$_3$ as a prototypical example. 
We then study monolayer BeO, a system featuring dark exciton state.
For both of them, we are able to explicitly calculate the mode- and momentum-resolved exciton-phonon coupling strength in the momentum space, from which we obtain the real-space electron (hole) localization. 
Using the phonon basis we can compute the lattice distortions and the STE potential energy surfaces.
Our method overcomes the difficulties concerning atomic force calculations in the presence of excitons and properly takes the many-body effects into account at a reasonable computational cost. 
It also allows us to explore a large configuration phase space, which is important to achieve realistic and accurate results. 
Finally, our approach can be easily applied to other semiconductors and insulators.

\textit{Theoretical framework.$-$}
We start from a perturbation-theory based framework, which has been applied to compute polaron formation~\cite{sio2019_polarons_prl,sio2019_abinitio_prb},  where the total energy of a system containing an extra electron can be described as a summation of its ground state energy and a linear expansion with respect to the lattice displacement.

\begin{eqnarray}
    \label{eq:dfpt}
    E\{\tau_{\kappa\alpha i},\psi\} &= E\{\tau^0_{\kappa\alpha i},\psi^0_{n\mathbf{k}}\} + \frac{1}{2}C_{\kappa_1 \alpha i, \kappa_2 \beta j}\tau_{\kappa_1 \alpha i}\tau_{\kappa_2 \beta j} \\ \nonumber
    &+ \int \psi^* [H^0_{KS}+\frac{\partial V^0_{KS}}{\partial \tau_{\kappa_1 \alpha i}} \tau_{\kappa_1 \alpha i}]\psi.
\end{eqnarray}
Here $\psi$ is the excess electron's wave function and $\tau_{\kappa\alpha i}$ is the atomic displacement with $\kappa\alpha i$ denoting the Cartesian coordinate $\alpha$ of atom $\kappa$ in the $i$th unit cell.
$H^0_{KS}$, $\frac{\partial V^0_{KS}}{\partial \tau_{\kappa_1 \alpha i}}$, and $\psi^0_{n\mathbf{k}}$ are the ground state Kohn-Sham (KS) Hamiltonian, the variation of the KS potential, and the wave function with band index $n$ and wave vector $\mathbf{k}$, respectively.
$C_{\kappa_1 \alpha i, \kappa_2 \beta j}$ represents the force constant matrix.

For an exciton-phonon coupled system, it is natural to extend the total energy expression by replacing the Kohn-Sham Hamiltonian with the many-body Bethe-Salpeter equation (BSE) Hamiltonian and the electronic wave function by its exciton counterpart:
\begin{eqnarray}
    E\{\tau_{\kappa\alpha i},\psi_{ex}\} &= E\{\tau^0_{\kappa\alpha i},\psi_{ex}^0\} + \frac{1}{2}C_{\kappa_1 \alpha i, \kappa_2 \beta j}\tau_{\kappa_1 \alpha i}\tau_{\kappa_2 \beta j} \\ \nonumber
    &+ \int \psi^*_{ex} [H_{BSE}+\frac{\partial H_{BSE}}{\partial \tau_{\kappa_1 \alpha i}} \tau_{\kappa_1 \alpha i}]\psi_{ex},
    \label{eq:exciton_dfpt}
\end{eqnarray}
where $\psi_{ex}^0$ denotes the exciton empty state, and the many-body BSE Hamiltonian can be written as
\begin{equation}
\label{eq:BSE_phonon_hamiltonian}
     H^{BSE}_{vc,v'c'}=(\varepsilon_c-\varepsilon_v)\delta_{cc'}\delta_{vv'}+(2V_{vc,v'c'}-W_{vc,v'c'}),
\end{equation}
with $W$ and $V$ representing the direct electron-electron attraction and the exchange term, respectively.
$c$ ($c'$) and $v$ ($v'$) are the conduction and valence band indices.

Although excitons satisfy the Bose commutation, we assume particle conservation can still be imposed under scenarios where excitons decay much slower than the STE formation.
Therefore, applying the principles of energy minimization and exciton number conservation to Eq.~\ref{eq:exciton_dfpt}, we reach the following set of self-consistent eigen equations:
\begin{eqnarray}\label{eq:self_consistent_exciton}
    &\frac{2}{N_p} B_{\mathbf{q}\mu}G^{\rm ex-ph}_{nm\mu}(\mathbf{Q},\mathbf{q})A_{m\mathbf{Q+q}}=(\varepsilon_{n\mathbf{Q}}-\varepsilon)A_{n\mathbf{Q}}, \\ \nonumber
    &B_{\mathbf{q}\mu} = \frac{1}{N_p} A^*_{m\mathbf{Q+q}}\frac{G^{\rm ex-ph}_{nm\mu}(\mathbf{Q,q})}{\hbar \omega_{\mathbf{q}\mu}}A_{n\mathbf{Q}}.
\end{eqnarray}
Here $N_p$ is the number of unit cells in the supercell, and $G^{\rm ex-ph}_{nm\mu}(\mathbf{Q,q})$ is the exciton-phonon matrix element. 
$A_{n\mathbf{Q}}$ denotes the STE wave function in the exciton basis, with $B_{\mathbf{q}\mu}$ the lattice wave function in the phonon eigenmode basis.
The coupling matrix of exciton and phonon is by definition the differential of the BSE Hamiltonian with respect to lattice displacement along the phonon normal mode.
Assuming the static screening function stays unchanged upon lattice distortion, one can estimate the exciton-phonon coupling matrix using the exciton wave function in the electron-hole pair basis $E^{n\mathbf{Q}}_{v\mathbf{k},c\mathbf{k+Q}}$ and electron-phonon coupling matrix $g^{\rm el-ph}$ by~\cite{antonius2022_theory_prb,chen2020_exciton_prl}
\begin{eqnarray}
    \label{eq:exciton_phonon_coupling_matrix}
    G^{\rm ex-ph}_{nm\mu}(\mathbf{Q,q})&=E^{m\mathbf{Q+q}*}_{v \mathbf{k},c \mathbf{k+Q+q}} E^{n\mathbf{Q}}_{v \mathbf{k},c' \mathbf{k+Q}}g^{\rm el-ph}_{c'c \mu}(\mathbf{k+Q,q}) \\ \nonumber
    &-E^{m\mathbf{Q+q}*}_{v \mathbf{k-q},c \mathbf{k+Q}} E^{n\mathbf{Q}}_{v' \mathbf{k},c \mathbf{k+Q}}g^{\rm el-ph}_{v v'\mu}(\mathbf{k-q,q}),
\end{eqnarray}
where $m$ and $n$ denote the exciton band indices and $\mathbf{Q}$ is the exciton center-of-mass momentum.

We would like to point out that, the above equations are obtained under the approximation that both phonons and the electron-phonon coupling lie in the linear regime, in other words under small lattice distortion.
Further, in practice, only a limited number of exciton bands will be included for the calculation to be affordable.
This requires well-defined exciton states, i.e. strong enough Coulomb interaction to maintain the exciton quasiparticle not broken down by the electron-phonon interaction.
We therefore first solve the BSE equation to obtain the lowest exciton bands, and then treat the exciton-phonon interaction perturbatively. 
This approximation holds well as long as the exciton binding energy is much larger than the STE formation energy. 

After solving the STE equations (Eq.~\ref{eq:self_consistent_exciton}), we obtain the STE wave function in the exciton basis $A_{n\mathbf{Q}}$, and the lattice wave function in the phonon basis $B_{\mathbf{q}\mu}$. 
The real space lattice distortion can be calculated by
\begin{equation}
    \tau_{\kappa \alpha i} = -\frac{2}{N_p}B^*_{\mathbf{q}\mu} \left(\frac{\hbar}{2M_\kappa \omega_{\mathbf{q}\mu}} \right)^{1/2}\mathbf{e}_{\kappa i \mu}(\mathbf{q})e^{i\mathbf{q}\cdot \mathbf{R}_i},
    \label{eq:tau}
\end{equation}
with $M_{\kappa}$ being the ion mass, $\mathbf{R}_i$ the direct lattice vector, $\mathbf{e}_{\kappa i \mu}$ the phonon normal mode with branch $\mu$, wave vector $\mathbf{q}$, and frequency $\omega_{\mathbf{q}\mu}$.

As such, our method offers another advantage to explore the potential energy surface of the free exciton and STE states by non-self-consistently solving the first eigen equation in Eq.~\ref{eq:self_consistent_exciton} with a phonon wave function that corresponds to a certain lattice distortion.
Otherwise, the formation energies can only be obtained from the finite displacement method using a large supercell, and suffer from a limited phonon phase space.

Furthermore, the ``localization'' of the STE state can be seen more clearly from the electron and hole distribution.
To this end, we obtain the STE wave function by further expanding the electron and hole Bloch wave functions following
\begin{equation}
    |\Psi_{STE} \rangle = \lambda_{vk_1}^{c k_2} |c k_2\rangle \langle vk_1|,
    \label{eq:ste wave function}
\end{equation}
with the coefficient $\lambda_{vk_1}^{c k_2} = A_{n\mathbf{Q}}E^{n\mathbf{Q}}_{vk_1,ck_2} $ ($\mathbf{Q}=k_2-k_1$). As can be seen from Eq.~\ref{eq:ste wave function}, the electron and hole states are entangled. 
We take advantage of the Wannier representation and calculate the reduced density matrix of the electron (hole) by tracing out the hole (electron) to reflect their real-space wave functions,

\begin{equation}
    \label{eq:rho hole}
    \hat{\rho}_{\rm hole}=\lambda_{v_1k_1}^{c k}\lambda_{v_2k_2}^{c k*} | v_2 k_2\rangle \langle v_1 k_1|,
\end{equation}
\begin{equation}
    \label{eq:rho electron}
    \hat{\rho}_{\rm electron}=\lambda_{vk}^{c_1 k_1}\lambda_{vk}^{c_2 k_2*} | c_1 k_1\rangle \langle c_2 k_2|,
\end{equation}
\begin{equation}
    \label{eq:rho wannier}
    \rho_{i w_1}^{j w_2}=\langle j w_2|\hat{\rho}_{\rm hole \ or \ electron}|i w_1 \rangle.
\end{equation}
Here, $i$ ($j$) denotes the unit cell index and $w$ denotes the Wannier function index.
The diagonal part of the reduced density matrix is the classical contribution to the electron (hole) density, and the off-diagonal part is the quantum coherence which leads to local density fluctuations. 
In the context of the electron (hole) distribution in the STE state, we only analyze the diagonal part for now.

\begin{figure*}[!ht]
    \centering
    \includegraphics[width=\linewidth]{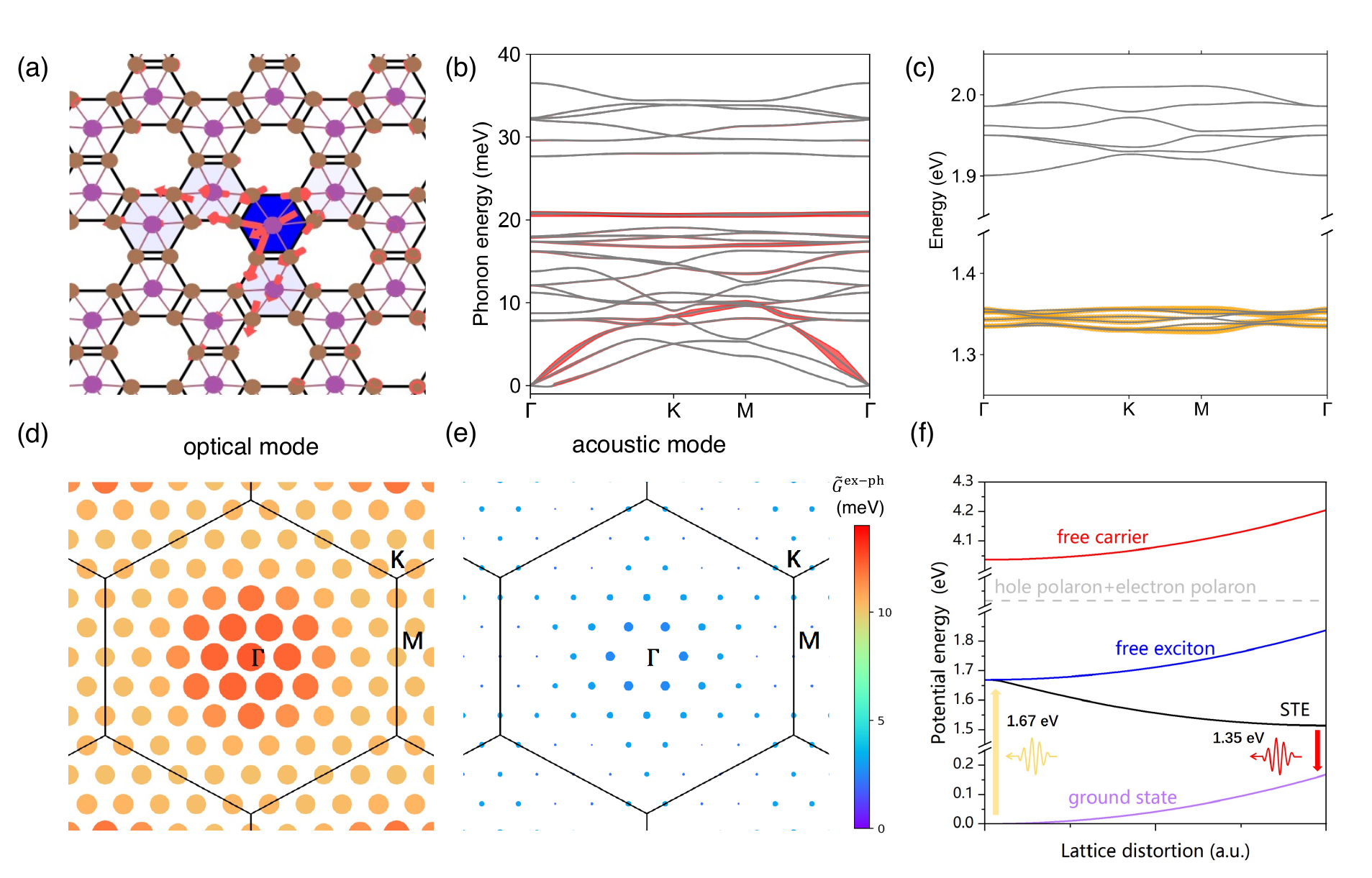}
    \caption{\textbf{STE in monolayer CrBr$_3$.} (a) The calculated atomic displacements on an $8\times8$ supercell of monolayer CrBr$_3$, labeled by red arrows. 
    Purple (brown) circles represent the Cr (Br) atoms, respectively.
    The blue colormap shows the hole density distribution, highlighting its real space localization. 
    (b) Phonon dispersion of monolayer CrBr$_3$ with its linewidth proportional to the amplitude of $|B_{\mathbf{q}\mu}|^2$.
    (c) Exciton bands of monolayer CrBr$_3$, with the radii of the circles proportional to the amplitude of $|A_{n\mathbf{Q}}|^2$.
    (d), (e) Distribution of the exciton-phonon coupling strength in the Brillouin zone for the flat optical and acoustic LA phonon modes, respectively.
    The color of markers indicates the value of $\tilde{G}^{\rm ex-ph} (\mathbf{q})$ while the size of the markers represents the phonon-frequency-normalized value$\frac{|\tilde{G}^{\rm ex-ph}_{\mu}(\mathbf{q})|^2}{\omega_{\mathbf{q}\mu}}$.
    (f) The potential energy surfaces (PESs) with respect to the lattice distortion. 
    The STE state (black curve) significantly lowers the total energy in comparison to the free exciton state (blue curve).
    Optical transition for the STE shows a red shift of 0.32~eV in comparison with the free exciton, shown by the bold arrows to guide the eye.}
    \label{fig:CrBr3-total}
\end{figure*}

\textit{STE in monolayer CrBr$_3$.$-$}Bulk CrBr$_3$ crystallizes in the trigonal $R\bar{3}m$ space group, and its monolayer contains edge-sharing CrBr$_6$ octahedra where Cr$^{3+}$ is bonded to six equivalent Br$^{1-}$ atoms. 
All Cr?Br bond lengths are 2.53~\AA. Br$^{1-}$  is bonded in an L-shaped geometry to two equivalent Cr$^{3+}$ atoms.
$G_0W_0$ calculation yields a band gap of 4.04~eV for monolayer CrBr$_3$, while BSE on top of $G_0W_0$ calculation gives the lowest exciton peak at 1.35~eV, which shows a slight red shift of $\sim0.32$~eV compared to the experimental result of 1.67~eV in bulk CrBr$_3$~\cite{MolinaSnchez2020_magneto_jmc}. 
The difference can also be possibly attributed to the absence of Hubbard U in our calculation, which has been shown to be responsible for a blue shift of around 400~meV for exciton peak in monolayer CrCl$_3$~\cite{zhu2020_quasiparticle_prb} (for detailed results of CrCl$_3$, see SI Sec.~S3). 
Meanwhile, the exciton bands in CrBr$_3$ are remarkably flat over the entire moment space, with a dispersion width of less than ten meV, as shown in Fig.~\ref{fig:CrBr3-total}(c).
Such flat exciton bands are a result of flat valance and conduction bands in CrBr$_3$, which could in turn facilitate the formation of STE by expanding the exciton-phonon phase space in a certain energy range (for the band structure of CrBr$_3$, see SI Fig.~S3).

Our calculation gives an STE formation energy of 156~meV in CrBr$_3$.
In combination with the lattice distortion energy of 168~meV, a total Stokes shift of 324~meV is expected, in excellent agreement with the experimental observation of 320~meV~\cite{Zhang2019_direct_nanolett}. 
Figure~\ref{fig:CrBr3-total}(f) explicitly shows the PES for the free carrier, the free exciton, and the STE states with respect to lattice distortion, with the electron (hole) polaron state marked by a dashed line for reference. 
One can see that the STE state is much more energetically favorable than the electron (hole) polaron state.
Another feature that can be observed is that the PES of the STE state has a significantly slower change at the energy minimum than the lattice distortion energy, meaning a small variation in STE energy can lead to a large variation of Stokes shift.
This naturally explains the large full width at half-maximum (FWHM) observed in the photoluminescence spectra~\cite{Zhang2019_direct_nanolett}.

Figures~\ref{fig:CrBr3-total}(b) and (c) show the phonon and exciton dispersion, on top of which we overlay the STE wave function in the exciton and phonon basis, to present the mode- and momentum-resolved contribution to the STE formation.
We notice that apart from the low energy longitudinal acoustic (LA) mode making a noticeable contribution to the STE formation, two almost degenerate flat phonon bands with frequency of $\sim20.58$~meV also have a strong coupling strength.
These two phonon modes correspond to the in-plane and out-of-plane breathing motion of the Cr-Br octahedra, respectively. 
It has been observed that similar phonon modes are excited coherently when the STE is formed in CrI$_3$~\cite{Li2022_ultrafast_nanolett}. 
The real space lattice distortion is shown by the red arrows in Fig.~\ref{fig:CrBr3-total}(a), each of which denotes our calculated atomic displacement by Eq.~\ref{eq:tau}.
The distortion is rather localized at the bound electron-hole pair, for which we show the hole distribution obtained from Eq.~\ref{eq:rho hole} in blue shading.
We further plot the distribution of averaged exciton-phonon coupling strength $\tilde{G}^{\rm ex-ph}$ for the phonon modes strongly-coupled to the exciton state of interest at $\mathbf{Q_0}$ in the entire Brillouin zone, which we define as
\begin{equation}
    \tilde{G}^{\rm ex-ph} (\mathbf{q}) = \frac{1}{N_b}\sqrt{\Sigma_{nm}|G^{\rm ex-ph}_{nm\mu}(\mathbf{Q_0},\mathbf{q})|^2}.
\end{equation}
Here $N_b$ is the number of the lowest exciton bands among which the average is taken.
As can be seen in Figs.~\ref{fig:CrBr3-total} (d) and (e), the flat optical mode exhibits larger and more uniform distribution coupling strength with the exciton states.
Together with the well localized exciton and lattice distortion, as well as the short-range nature of exciton-phonon interaction, we believe our method successfully captures the STE formation in monolayer CrBr$_3$.

\begin{figure*}[!ht]
    \centering
    \includegraphics[width=\linewidth]{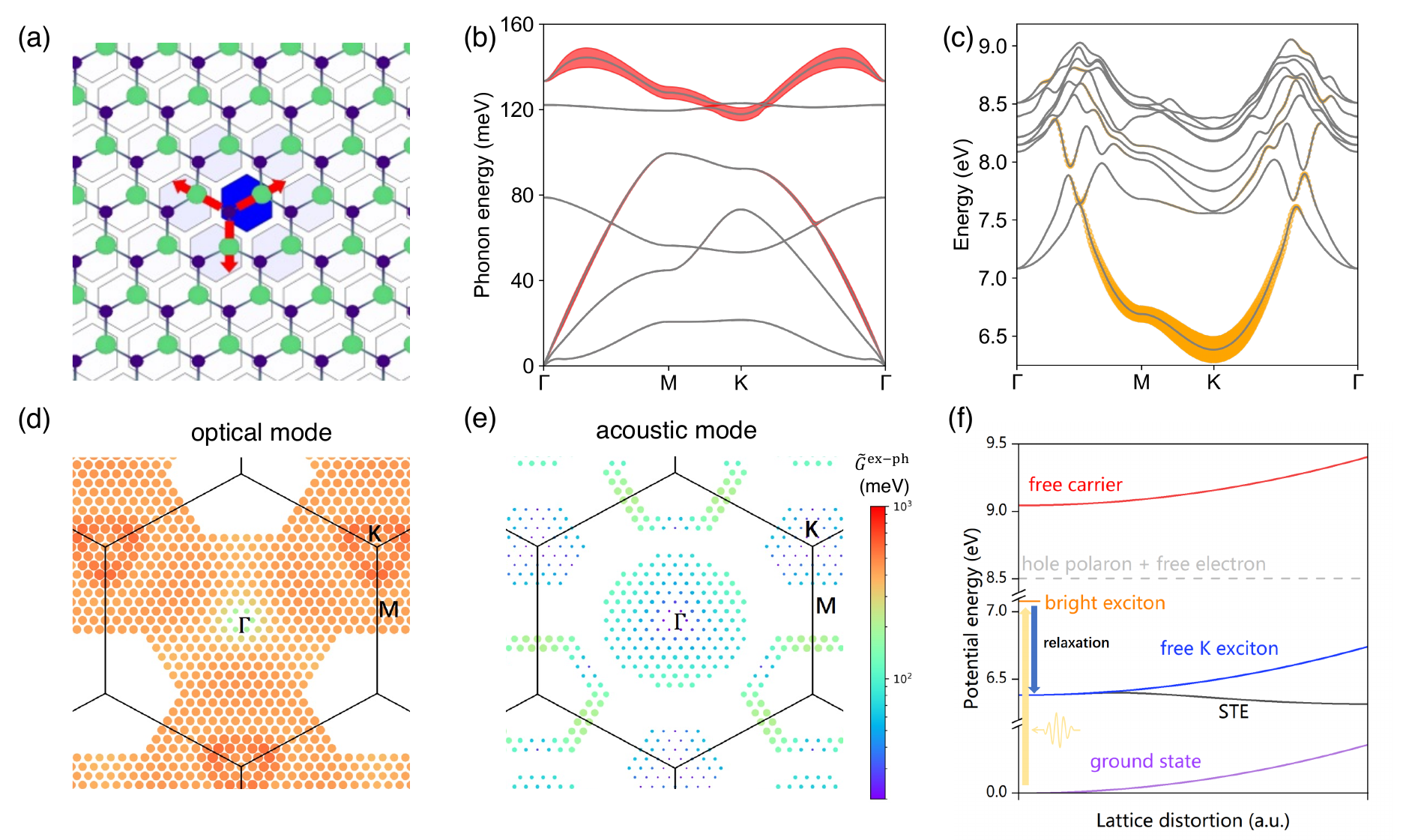}
    \caption{
    \textbf{STE in monolayer BeO.} (a) The calculated atomic displacements on a $24\times24$ supercell of monolayer BeO, labeled by red arrows. 
    Green (dark blue) circles represent the Be (O) atoms, respectively.
    The blue colormap shows the hole density distribution, highlighting its real space localization. 
    (b) Phonon dispersion of monolayer BeO with its linewidth proportional to the amplitude of $|B_{\mathbf{q}\mu}|^2$.
    (c) Exciton bands of monolayer BeO, with the radii of the circles proportional to the amplitude of $|A_{n\mathbf{Q}}|^2$.
    Notice the exciton bands in BeO is highly dispersed with the global minimum located at $K$.
    (d), (e) Distribution of the exciton-phonon coupling strength in the Brillouin zone for the optical and acoustic phonon modes, respectively.
    The color of markers again indicates the value of $\tilde{G}^{\rm ex-ph} (\mathbf{q})$ while the size of the markers represents the phonon-frequency-normalized value $\frac{|\tilde{G}^{\rm ex-ph}_{\mu}(\mathbf{q})|^2}{\omega_{\mathbf{q}\mu}}$.
    (f) The potential energy surfaces (PESs) with respect to the lattice distortion. 
    The STE state (black curve) lowers the total energy in comparison to the free exciton state (blue curve).
    The grey dashed line shows the energy of hole polaron with an excess electron, significantly higher than the STE state.
    The bold arrows highlight the optical transition of the bright exciton at 7.08~eV before relaxing to the dark exciton through coupling to phonons.}
    \label{fig:BeO-total}
\end{figure*}

\textit{STE in monolayer BeO.$-$}Recently, monolayer BeO was synthesized through the epitaxial growth method~\cite{Zhang_2021_monolayerBeO}, which broadens the 2D materials phase space beyond being limited to their existing bulk layered counterparts.
BeO has a honeycomb lattice similar to that of hexagonal boron nitride (hBN). 
As an insulator, it has a wide band gap ($>5$~eV), suggesting it is likely to exhibit a significant excitonic effect. 
Moreover, a recent work reports a localized hole polaron with formation energy of over 500~meV in 2D BeO, indicating strong electron-phonon coupling~\cite{Sio_2023_2D_polaron}.
These facts motivate us to study the STE formation and make predictions that we hope can be verified by future experiments.

DFT level calculation shows a direct gap of 5.85~eV and an indirect gap of 5.36~eV in monolayer BeO.
$G_0W_0$ calculation yields an overall band gap correction of around 3.7~eV. 
BSE calculation results in the lowest bright exciton at 7.08~eV with a binding energy of 2.5~eV. 
Such a large binding energy implies the lowest energy exciton in monolayer BeO most likely has a Frenkel-type nature.
Using the same approach, we obtain an STE formation energy of 66~meV, with a lattice distortion energy of 360~meV. 
Compared to the hole polaron formation energy of 542~meV obtained in this work, we find the STE is $\sim2$~eV more stable than a hole polaron and a free electron located at the conduction band minimum. 
Thus, the STE is unlikely to be broken down by strong interaction with the lattice, therefore monolayer BeO may serve as an ideal testbed for STE formation.

From the phonon band plot in Fig.~\ref{fig:BeO-total}(b), we can see that both LA and LO phonons play a major role in the STE formation in monolayer BeO, while the remaining phonon branches make a negligible contribution. 
This is similar to the phonon contribution to the hole polaron formation analyzed in the previous work~\cite{Sio_2023_2D_polaron}.
However, there lies a major difference.
In the hole polaron, the long wavelength LO phonon dominates through long-range electron-phonon coupling~\cite{Sio_2023_2D_polaron}.
Excitons, however, as a charge neutral quasiparticle, can interact with phonons not limited to the long wavelength modes.
Thus, the contribution of different wave vectors of the LO phonon branch is almost evenly distributed, which is further illustrated in Figs.~\ref{fig:BeO-total}(d) and (e).
There, we plot the averaged exciton-phonon coupling strength of the $K$-valley exciton, due to the indirect-gap nature of the system.
One can see that short wavelength phonons with a specific momenta show much stronger coupling compared to that at the $\Gamma$ point, in line with the physical picture of exciton-phonon coupling. 

Further, when the STE is formed in monolayer BeO, the extra hole is mainly located at the O atom where the lattice distortion is centered on, while the electron is located on the neighboring Be atom.
The real space distortion is also computed and shown in Fig.~\ref{fig:BeO-total}(a), where the Be atoms are pushed away from the O atom, i.e. the localized hole, leading to a bond length changing from 1.54~\AA~to 1.63~\AA.
Another noticeable feature is that the hole seems to be more localized than the electron when the STE is formed (see SI Fig.~S4 for comparison). 
This is consistent with the existence of small hole polaron and absence of electron polaron in monolayer BeO, implying the lattice distortion mainly traps the exciton through strong lattice-hole interaction, and the electron is trapped through strong attractive Coulomb interaction between the electron and the trapped hole.

Finally, we note that monolayer BeO is an indirect gap semiconductor (for band structure of BeO, see SI, Fig.~S3), where the lowest exciton state is located at the $K$ valley and has an energy 0.7~eV lower than the lowest bright exciton at the $\Gamma$ point.
In the same spirit of previous calculations, we may expect a huge Stokes shift of $\sim 1.1$~eV. 
However, due to the indirect gap nature of monolayer BeO, its STE is mainly composed of excitons localized at the $K$ valley which is also the global minimum for the exciton bands (Fig.~\ref{fig:BeO-total}(c)).
Therefore, STE in BeO has a much lower tendency to decay radiatively due to momentum mismatch, thus how it affects the photoluminescence will need further study.

\textit{Conclusions.$-$}
In conclusion, we propose a theoretical and computational framework to treat the self-trapped excitons from first principles. 
Our method properly takes into account the electron-hole many-body interaction and their coupling to the full lattice degree of freedom using a combination of the Bethe-Salpeter equation and perturbation theory. 
Using a two-dimensional magnetic semiconductor CrBr$_3$ and a wide-gap indirect gap semiconductor BeO as prototype examples, we are able to quantitatively discuss the fine structure of the STE potential energy surfaces and compare them to the polaronic states.
We also evaluate their Stokes shift energies that could readily be verified by the photoluminescence experiments.
From the mode- and momentum-resolved coupling strength calculation, we pinpoint the phonon modes with significant contributions that could potentially be excited coherently and observed in transient absorption spectroscopy.
Our work highlights the importance of lattice degree of freedom as well as the many-body effects, and further provides a general framework to effectively and efficiently make predictions of STE in real materials from first principles.
Finally, the efficiency of our scheme can be significantly improved if the exciton-phonon coupling matrix can be interpolated using appropriate schemes, such as those employed for calculating the electron-phonon systems.

\textit{Acknowledgements.$-$}The authors thank the Ministry of Science and Technology (No.~2021YFA1400201), National Natural Science Foundation of China (Nos.~12025407, 11934004, and 92250303), and Chinese Academy of Sciences (Nos.~YSBR-047 and XDB33030100) for financial support.

\bibliography{main_ref}
\end{document}